\documentclass[12pt]{iopart}

\usepackage{graphicx}
\usepackage{amssymb}

\begin{document}

\title{Electron flow in split-gated bilayer graphene}

\author{Susanne Dr\"oscher$^{1}$, Cl\'ement Barraud$^{1}$, Kenji Watanabe$^2$, Takashi Taniguchi$^2$, Thomas Ihn$^1$ and Klaus Ensslin$^1$}


\address{$^1$ Solid State Physics Laboratory, ETH Zurich, 8093 Zurich, Switzerland}
\address{$^2$ Advanced Materials Laboratory, National Institute for Materials Science, 1-1 Namiki, Tsukuba, 305-0044, Japan}

\ead{susanned@phys.ethz.ch}

\begin{abstract}

We present transport measurements on a bilayer graphene sheet with homogeneous back gate and split top gate. The electronic transport data indicates the capability to direct electron flow through graphene nanostructures purely defined by electrostatic gating. By comparing the transconductance data recorded for different top gate geometries - continuous barrier and split-gate - the observed transport features for the split-gate can be attributed to interference effects inside the narrow opening.
\end{abstract}

\maketitle

\section{Introduction}
Graphene nanostructures have triggered an intense effort of research both theoretically and experimentally. Proposals for the manipulation of individual spin states in graphene quantum dots have been developed \cite{Trauzettel:2007} and excited charge and spin states in graphene quantum dots have been experimentally observed \cite{Guettinger:2010}. These results show that confining single electrons is possible and constitute an important step towards fully controlled electronic states in graphene nanostructures.

Up to now, nanostructures in graphene are mainly realized by etching the flakes through a lithographically defined mask \cite{Molitor:2011, Droscher:2011}. Split-gate defined nanostructures cannot be realized in single layer graphene because an intrinsic band gap is absent and Klein tunneling occurs between n- and p-doped regions \cite{Stander:2009}, which prevents electrostatic gating. A promising technological alternative has been unveiled by the use of bilayer graphene since a band gap can be electrically induced and tuned in this material \cite{Oostinga:2008, Russo:2009}. Electrostatic confinement of carriers thus becomes possible similar to semiconductor-based devices \cite{Wees:1988}. The apparent advantage of this method is the smoother confinement potential expected in contrast to rough and uncontrolled edges on the microscopic scale in present graphene nano-devices.

In this paper, we show that the flow of charge carriers can indeed be controlled and directed through nanometer sized constrictions by electrostatic means only. We have measured mesoscopic fluctuations of the gate voltage dependent conductance, which are interpreted as interference effects from charge carriers being transferred through the narrow channel defined by the split-gates. The results suggest that this device is in a different transport regime than those recently reported in Refs. \cite{Allen:2012} and \cite{Goossens:2012}.

\section{Sample and measurement setup}

\begin{figure}[t!]
 \begin{flushright}
    \includegraphics[width=13cm]{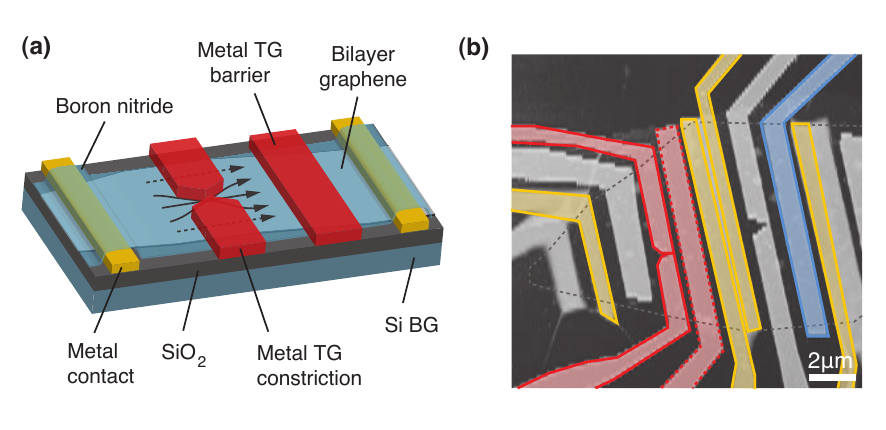}
    \caption{(a) Schematic of a double gated bilayer graphene flake. The graphene flake resides on Si/SiO$_2$ substrate and is electrically contacted with metallic electrodes (yellow). The boron nitride flake (light blue) covers the graphene and these electrodes. Top gate fingers (red and blue) are deposited last. The gray arrows indicate, where current can flow through the split-gate structure. (b) AFM image of the bilayer graphene device. The black dotted line outlines the graphene flake, the red solid line indicates the split-gate structure considered in this paper and the red dotted line outlines the barrier, which is directly compared to the constriction. The blue solid line marks the barrier measured separately and presented in Fig.~\ref{c8fig2} and the yellow solid lines mark the Ohmic contacts used. The width of all top gate fingers is $W$ = 900 nm.}
    \label{c8fig1}
  \end{flushright}
\end{figure}

Mechanical exfoliation of natural graphite is used to extract graphene flakes and deposit them onto a Si/SiO$_2$ substrate \cite{Novoselov:2004}. Thin flakes were identified using an optical microscope. Subsequent atomic force microscope (AFM) measurements as well as Raman spectroscopy \cite{Ferrari:2006,Graf:2007} were performed to verify the flakes' bilayer nature. Ohmic contacts were defined by electron beam lithography (EBL) followed by metal evaporation (2~nm Cr and 40~nm Au) and a lift-off process. Hexagonal boron nitride (BN) was deposited on the bilayer flake as a dielectric material. Since micrometer-sized flakes of this material are mechanically exfoliated as well, we used the mechanical transfer technique developed by Dean et al. \cite{Dean:2010} to place a BN flake at the desired position on a chip. After covering the contacted graphene flake with $\approx$ 7.5~nm thick BN, top gate electrode patterns were defined by EBL. Finally, metal was evaporated (0.5~nm Cr and 45~nm Au) and the gate fingers were revealed by a subsequent lift-off. A schematic of such a sample is shown in Fig.~\ref{c8fig1}~(a). Panel (b) of the same figure displays an atomic force micrograph of the device studied here.

With standard lock-in techniques we recorded the three-terminal voltage drop induced by a constant current bias ($I_\mathrm{bias}$ = 0.5 nA). Since the outer left contact in Fig.~\ref{c8fig1}~(b) was not working, a conventional four-terminal measurement was not possible and the contact marked yellow on the left hand side was used for both current injection and voltage probing. 
Besides, we measured the change of resistance by modulating the gates with a small AC signal with frequency $f <$ 100 Hz superimposed onto their DC bias voltage $V_\mathrm{TG}$. With a lock-in amplifier we then detected the modulated three-terminal voltage at the same frequency $f$ and obtained the transconductance signal from $\partial G/\partial V_\mathrm{TG}=R_\mathrm{DC}^{-2}\cdot\partial R/\partial V_\mathrm{TG}$. If not stated otherwise, the measurements were carried out in a variable temperature insert at a base temperature of $T\approx$ 1.7~K.

\section{Results and discussion}

\subsection{Comparison of different geometries}

\begin{figure}[t!]
 \begin{flushright}
    \includegraphics[width=13cm]{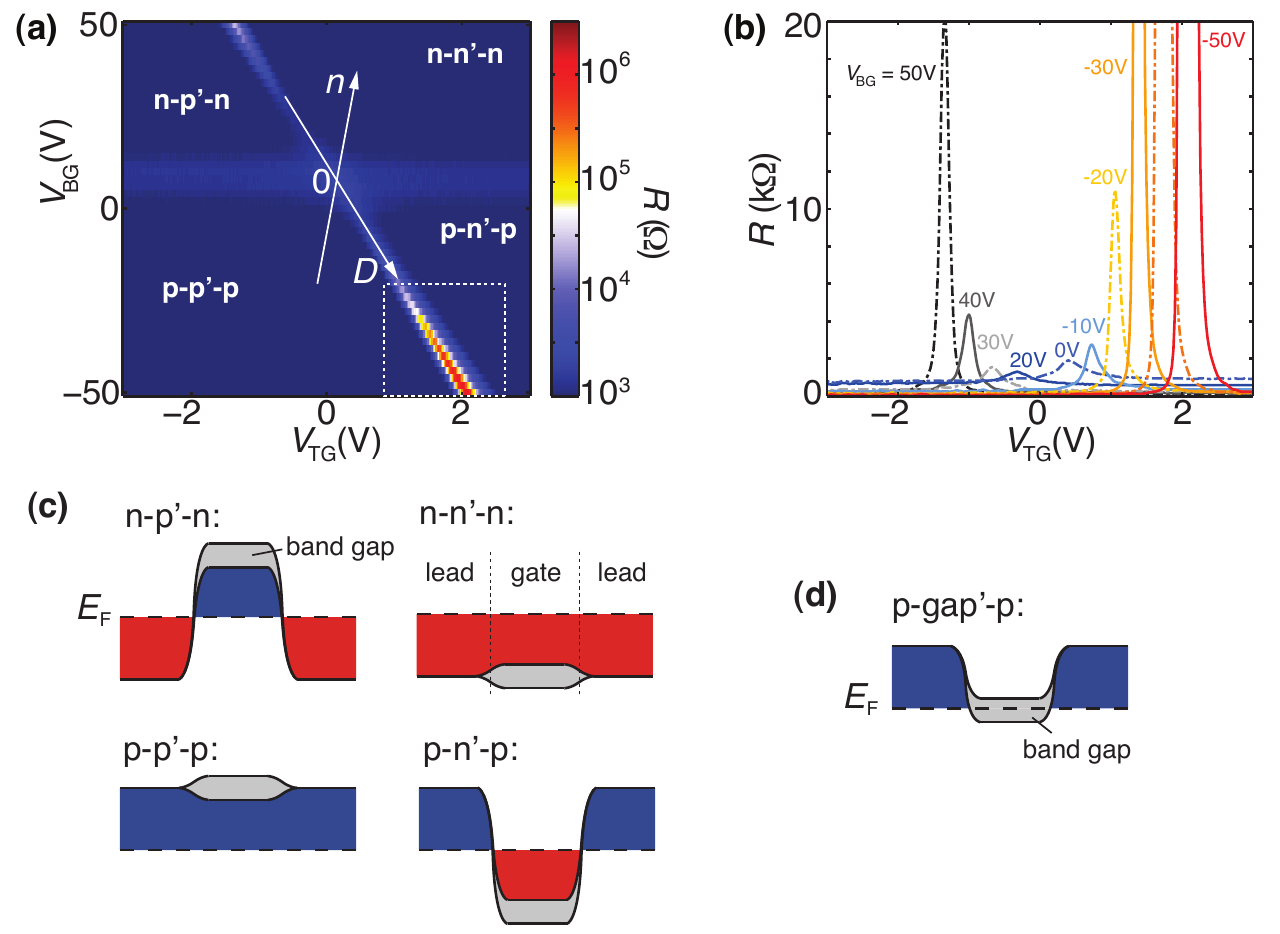}
    \caption{(a) Four point resistance as a function of top gate and back gate for the barrier marked blue in Fig.~\ref{c8fig1}~(b). The dashed rectangle indicates the gate space considered in the measurements below. (b) $R(V_\mathrm{TG})$ at constant $V_\mathrm{BG}$ between -50 V and +50 V. The back gate was changed in steps of 10 V and the current bias was $I_\mathrm{bias}$ = 1 nA. The four-terminal voltage drop was measured. (c) Sketch of the electronic band alignment for the four gate configurations identified in (a). (d) Band alignment for the transition from p-p'-p to p-n'-p. In the vicinity of $n$ = 0 a band gap is present underneath the gate.}
    \label{c8fig2}
  \end{flushright}
\end{figure}

First, we verified that a band gap can be opened in the region of the bilayer graphene flake covered by the continuous top gate stripe marked with the blue solid line in Fig.~\ref{c8fig1}~(b). In order to split the valence and conduction band edges in bilayer graphene, the layer symmetry has to be broken. Such an asymmetry is imposed by an electric field oriented perpendicularly to the bilayer graphene plane \cite{McCann:2006}. In Fig.~\ref{c8fig2} measurements of the resistance through the device as a function of back gate and top gate voltage are shown. The arrows shown in Fig.~\ref{c8fig2}~(a) indicate the axes along which the density $n$ in the graphene and the displacement field $D$ across the layers are tuned in the top-gated area of the device. These two parameters are defined by the applied gate voltages via the field effect. Moving along the displacement axis at zero density, the maximal achievable values are $D$ = -1.64 V/nm for the negative $V_\mathrm{BG}$ regime and $D$ =  1.05 V/nm for positive values of $V_\mathrm{BG}$. The transmission below the top gate was significantly reduced at high layer asymmetry as seen in Fig.~\ref{c8fig2}~(b). More specifically, the resistance at the Dirac point could be increased from 1.6~k$\Omega$ at $D$ = 0 to 1.6~M$\Omega$ at $D$ = -1.64~V/nm. This resistance increase by a factor of 1000 indicates the appearance of an insulating state underneath the top gate. The analysis of the temperature dependence of the resistance shows, that subgap states are present \cite{Shklovskii:1984} that lead to an enhanced conductance at the lowest temperature compared to the expected exponential decay of the conductance in case of a defect free gap. Our experimental results agree with those of Refs. \cite{Oostinga:2008} and \cite{Russo:2009}, where a variable-range hopping mechanism through localized states in the gap was proposed to explain the temperature dependence.

For the further discussion we would like to identify four distinct gate configurations in Fig.~\ref{c8fig2}~(a) leading to different alignment of the electronic bands as schematically shown in Fig.~\ref{c8fig2}~(c). The displacement axis separates the region of p-type doping (labeled p') below the top gate on the left hand side from that of n-type doping (labeled n') underneath the TG. The areas of the graphene that are not covered by a top gate electrode contribute to transport as well. They are tuned by the back gate only. The horizontal stripe of increased resistance around $V_\mathrm{BG}$ = 10 V displays the charge neutrality point in these areas. Hence, changing the back gate voltage from negative to positive values induces a transition from the valence band (labeled p) to the conduction band (labeled n). Depending on the specific configuration either unipolar (p-p'-p and n-n'-n) or bipolar (n-p'-n and p-n'-p) junctions can be formed between the adjacent graphene areas.

In the following, we present a comparison of the two different structures marked with the red outlines in Fig.~\ref{c8fig1}~(b). Since the resistance was increased by a factor of five upon opening the gap for the barrier considered here (red dotted line), current flow is never completely suppressed underneath the gate. This finding reflects the increased disorder in this part of the flake as compared to the part discussed previously on the basis of Fig.~\ref{c8fig2}. The neighboring constriction is likely to exhibit similar properties and transport happens both through the narrow channel and below the gates as indicated in  Fig.~\ref{c8fig1}~(a). In order to separate effects that stem from the ungated area from those that originate from the gated regions, we chose to measure transconductance for these structures by modulating the gate voltages. We wish to find out, to what extent charge carriers can be forced to flow through the narrow constriction of the split gate. For both devices we looked at a zoom into the negative displacement range of the two-dimensional conductance map as marked in Fig.~\ref{c8fig2}~(a), where we expect the transition from a unipolar (p-p'-p) to a bipolar (p-n'-p) junction as we change the carriers underneath the TGs from holes (p') to electrons (n'). As depicted in Fig.~\ref{c8fig2}~(d), a band gap is expected to be present underneath the TGs for the gate configuration at this transition. 

\begin{figure}[b!]
 \begin{flushright}
    \includegraphics[width=13cm]{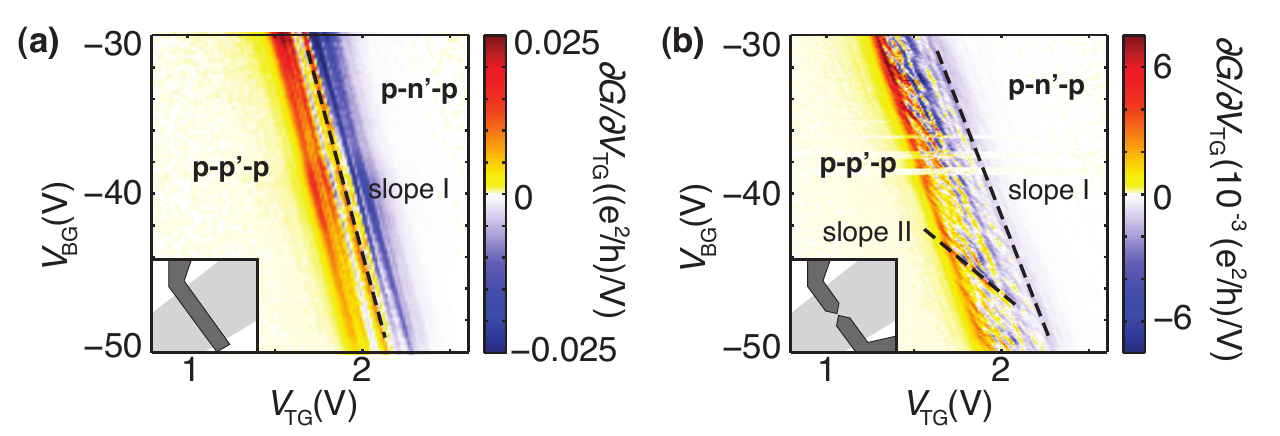}
    \caption{Transconductance maps for (a) full barrier (dotted outline in Fig.~\ref{c8fig1}~(a)) and (b) split-gates (solid outline in Fig.~\ref{c8fig1}~(a)). The sketches illustrate the respective geometry. The dashed lines are a guide to the eye for the two slopes discussed in the main text. The measurements were carried out in a dilution fridge at $T$ = 100~mK. The current bias was 0.5~nA for all experiments and the modulation amplitude for the transconductance was $\Delta V_\mathrm{AC}$ = 10~mV.}
    \label{c8fig3}
  \end{flushright}
\end{figure}

We start with the discussion of the barrier gate that completely spans the width of the graphene flake. In the transconductance plot in Fig.~\ref{c8fig3}~(a) the region of suppressed transmission displays as a stripe of large relative resistance fluctuations. A number of resonance minima and maxima running parallel to the displacement axis are visible (labeled 'slope I'). We attribute these oscillations to localized states located underneath the top gate, due to their relative lever arm in gate space.

The transport characteristics are different in the part of the sample where a narrow gap ($W_\mathrm{ch}$ = 80~nm) is present between two gates, as shown in Fig.~\ref{c8fig3}~(b). Again, the transconductance signal shows oscillations resulting in local minima and maxima running parallel to the displacement axis (slope I). This indicates that localized states in the top-gated areas contribute to transport again. Strikingly, an additional set of resonances appears, which exhibits a considerably lower relative lever arm $\alpha_\mathrm{TG}/\alpha_\mathrm{BG}$ and is hence less strongly tuned by the top gates (slope II). Both slopes are marked in Fig.~\ref{c8fig3}~(b) with dashed lines. Similar slope II resonances were observed in more than five different split-gate devices in total but never in devices with an ungapped gate. Therefore they originate from the narrow opening. As the displacement field is lowered, first the newly observed slope II resonances fade away at $D\approx$ $-$0.73~V/nm (not shown here), whereas the slope I resonances remain visible down to $D\approx$ $-$0.4~V/nm. This finding indicates, that the charge carriers are more and more forced to pass through the channel below the gap of the split-gate as $D$ is increased even though the confinement is not perfect. We will discuss the origin of these oscillations below.

We conclude that the gate structure affects the transmission and that charge carriers tend to be confined to a narrow channel by split-gates in the presence of a perpendicular electric field. This is the main result of the paper. In the following section we focus in more detail on the set of oscillations, which we attribute to the opening between the split-gates (slope II) and aim to understand their origin.

\subsection{Possible transport mechanisms}

A number of mechanisms are conceivable to cause the observed oscillations of different slopes in the transport data. Although transport through discrete modes leading to quantized conductance are expected for an ideal nanoconstriction, we will also consider alternative explanations. Namely, resonant tunneling or Coulomb blockade through states inside the channel and universal conductance fluctuations in the leads would induce an oscillatory behavior as well.

\begin{figure}[b!]
 \begin{flushright}
    \includegraphics[width=13cm]{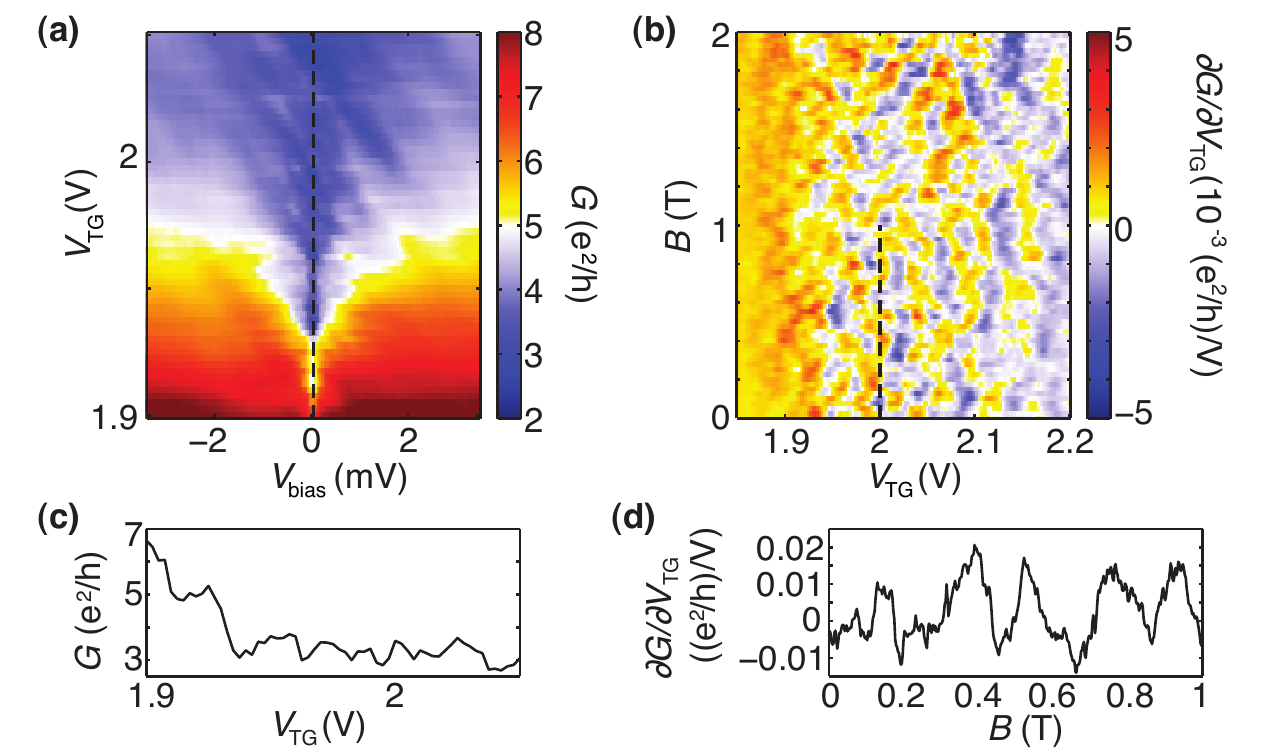}
    \caption{(a) Finite-bias spectroscopy as a function of applied top gate voltage $V_\mathrm{TG}$ taken at $V_\mathrm{BG}$ = $-$50~V carried out inside the region of suppressed conductance (see dashed line in Fig.~\ref{c8fig4}~(a)). (c) Cut at $V_\mathrm{bias}$ = 0 in (a). The conductance measurement was taken at $T$ = 100~mK with standard lock-in techniques using an AC modulation voltage of 50~$\mu$V. (b) Transconductance map showing the $B$-field dependence of a TG trace taken at $V_\mathrm{BG}$ = $-$50~V. (d) Cut in (b) at the position indicated by the dashed line. The measurements were performed at $T$ = 100~mK with a current bias of 0.5~nA and the modulation amplitude for the transconductance was $\Delta V_\mathrm{AC}$ = 10~mV.}
    \label{c8fig6}
  \end{flushright}
\end{figure}

Transport via discrete modes can be excluded since the DC data (see e.g. Fig.~\ref{c8fig6}~(c)) does not show quantized steps and the estimated mean free path of $l_\mathrm{mfp}\lesssim$ 100~nm in the density range where the resonances appear, indicates that the system is not fully ballistic. Further, resonant tunneling through localized states or Coulomb blockade inside the constriction cannot explain the experimental data because, in order to observe these effects, the resistance of the tunneling barriers is required to exceed the resistance quantum $R_\mathrm{T}=h/e^2$. As apparent in the conductance trace displayed in Fig.~\ref{c8fig6}~(c), the conductance never falls below 2$e^2/h$ and in the finite-bias spectroscopy measurement shown in Fig.~\ref{c8fig6}~(a) Coulomb blockade diamonds cannot be distinguished.

Universal conductance fluctuations (UCF) occur in open diffusive systems with conductances larger than the conductance quantum \cite{Beenakker:1991}. They can be viewed as originating from quantum interference of elastically scattered waves propagating along different paths, leading to a reflection amplitude fluctuating with carrier density by $\Delta G = e^2/h$ at zero temperature \cite{Lee:1985}. UCF may be the cause of the oscillations we observe in the data of Fig.~\ref{c8fig3}~(b) and Fig.~\ref{c8fig4} (to be discussed in detail below).

Since interference effects are altered by an external perpendicular magnetic field, a magnetic correlation field can be infered from measurements of the conductance as a function of $B$ \cite{Lee:1987}. Figure \ref{c8fig6}~(b) shows the magnetic field dependence of the transconductance measured along the same gate configuration  as Fig.~\ref{c8fig6}~(a). The oscillations exhibit a quasiperiodic pattern typical for conductance fluctuations as a function of magnetic field (see Fig.~\ref{c8fig6}~(d)). From the value of this mean periodicity, $\Delta B \approx$ 200~mT, we can estimate the area covered by interfering paths to be $A = \left(h/e\right)/\Delta B\approx$ 150~nm$^2$ comparable to the area of the constriction. This is a lower bound for the covered area, since the correlation field is expected to be lower than the here determined $\Delta B$. As can be seen in Fig.~\ref{c8fig6}~(c), the amplitude of the conductance oscillation is $\Delta G \approx e^2/h$, which further supports the presence of universal conductance fluctuations in the system.

The above considerations show that the system is not in the ballistic regime and hence quantized conductance is not expected. Further, tunneling effects can be excluded due to the low resistance of the device. It is however most likely that we probe universal conductance fluctuations of an area around the opening between the gates. Since the mean free path is on the order of the width and length of the opening, the characteristics of the present device are located at the cross-over from ballistic to diffusive transport. Additionally, we cross over from an unconfined to a confined system as the band gap is opened.

\subsection{Correlations between the transconductance of individual gates}

In the following we suggest a simple model to explain the observed transport in the present device in terms of conductances. The probed region of the structure consists of a conducting channel through the narrow constriction, which contributes $G_\mathrm{constr}$ to the transport. Further the conductance underneath the upper top gate $G_\mathrm{up}$ and the conductance below the lower top gate $G_\mathrm{low}$ need to be considered. These three conductances are connected in parallel and the total conductance is hence given by their sum
\begin{eqnarray}
 G_\mathrm{tot}(V_\mathrm{TG,up},V_\mathrm{TG,low},V_\mathrm{BG}) \nonumber \\
 = G_\mathrm{up}(V_\mathrm{TG,up},V_\mathrm{BG})+G_\mathrm{low}(V_\mathrm{TG,low},V_\mathrm{BG})+G_\mathrm{constr}(V_\mathrm{TG,up},V_\mathrm{TG,low},V_\mathrm{BG}).
 \label{Gtot}
\end{eqnarray}

So far, the transconductance signal displayed was obtained by modulating both top gates simultaneously. The recorded data shows $\partial G_\mathrm{tot}/\partial V_\mathrm{TG}$. We expect the different terms in Eq.~(\ref{Gtot}) to cause the two observed sets of oscillations with different slopes discussed in Fig.~\ref{c8fig3}~(b). Whereas the slope I oscillations stem from the first two terms on the right hand side, the slope II resonances are caused by the constriction, which is represented by the last term on the right hand side of Eq.~(\ref{Gtot}). 

\begin{figure}[ht]
 \begin{flushright}
    \includegraphics[width=13cm]{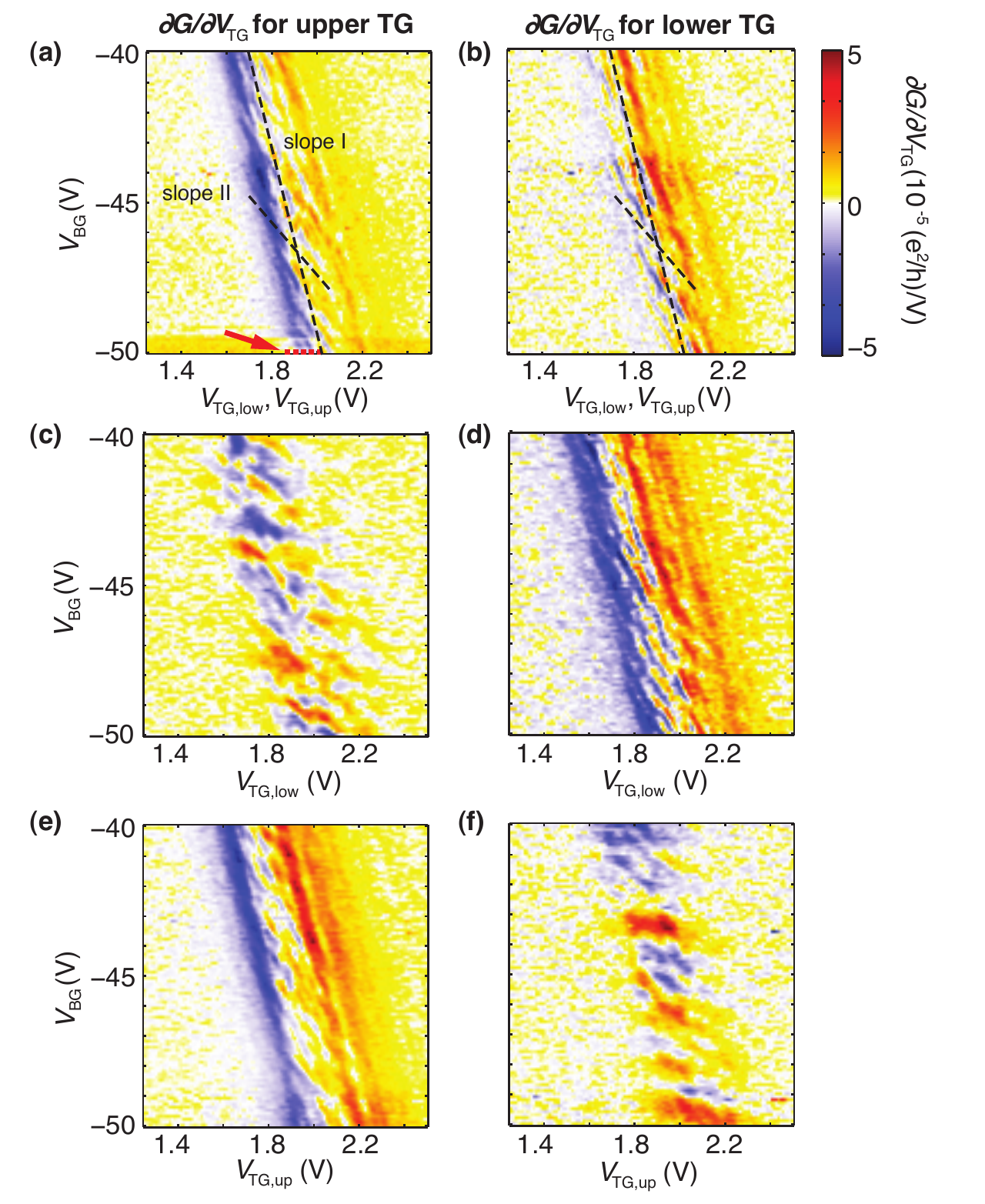}
    \caption{Transconductance maps for the individual TGs. Left (right) column displays the signal obtained by modulating the upper (lower) TG. (a) and (b): The DC voltages on both gates are swept simultaneously. The arrow in (a) indicates the position of the conductance traces of Fig.~\ref{c8fig6}~(c) and the black dashed lines in (a) and (b) show the line along which the TG is tuned for the measurements displayed in the 2$^\mathrm{nd}$ and 3$^\mathrm{rd}$ row. (c) and (d): DC voltage for upper TG kept at charge neutrality ($n$ = 0), DC voltage for lower TG swept. (e) and (f) Same as second column but vice versa. The current bias was 0.5~nA for all experiments and the modulation amplitude for the transconductance was $\Delta V_\mathrm{AC}$ = 10~mV. Colorbar is valid for all plots.}
    \label{c8fig4}
  \end{flushright}
\end{figure}

To test this model further and to understand the contribution to the total transconductance signal made by each of the two top gates, we measured their transconductances individually. For these measurements we superimposed a modulation voltage $\Delta V_\mathrm{AC}$ = 10~mV onto the DC voltage applied to each of the gates at a different frequency ($f_\mathrm{up}$ = 71.0~Hz and  $f_\mathrm{low}$ = 13.3~Hz) and detected the transconductance signal $\partial G/\partial V_\mathrm{TG}$ at the respective frequency. The results are shown in Fig.~\ref{c8fig4}, where the left column displays the transconductance signal recorded for the upper TG (at $f_\mathrm{up}$) and the right column corresponds to the data recorded for the lower TG (at $f_\mathrm{low}$).

In the measurements leading to the first row of Fig.~\ref{c8fig4}, the DC voltages applied to the gates were swept simultaneously between 1.2~V and 2.5~V. The transconductance maps show similar features to the one observed in Fig.~\ref{c8fig3}~(b), namely two sets of resonances - one with slope I and another one with slope II. The fact that these latter oscillations appear when we modulate any of the two gates, indicates that their physical origin lies in the vicinity of both gates. Since the relative lever arms are the same for the upper and the lower gate, we conclude that the capacitive coupling is equally strong and hence the states tuned have similar distance from both gates.

We now would like to verify further, that we can distinguish the three conductance channels introduced in Eq.~(\ref{Gtot}). For the data displayed in the second row of Fig.~\ref{c8fig4}, only the lower top gate voltage was swept between 1.2~V and 2.5~V. The DC voltage for the upper gate was swept as a function of the back gate voltage according to $V_\mathrm{TG,up}= \alpha_\mathrm{TG,up} \cdot V_\mathrm{BG}$ and in doing so the charge carrier density below this top gate was held at the position of the CNP (see dotted line in Fig.~\ref{c8fig4}~(a)). This means that a band gap is expected underneath the upper top gate throughout this measurement leading to $G_\mathrm{up}=G_\mathrm{up,min}$ and the remaining two conductances $G_\mathrm{low}$ and  $G_\mathrm{constr}$ dominate transport under these conditions.

In Fig.~\ref{c8fig4}~(c) the transconductance signal recorded for the upper TG at $f_\mathrm{up}$ is shown. A number of resonances with slope II are visible but the slope I oscillations are absent here. Considering the above model and the fact that the conductance $G_\mathrm{up}$ is strongly suppressed, the total transconductance is given by $\partial G_\mathrm{tot}/\partial V_\mathrm{TG,up}\approx\partial G_\mathrm{constr}/\partial V_\mathrm{TG,up}$. In this case, mainly fluctuations with slope II are detected.

For the data displayed in Fig.~\ref{c8fig4}~(d) the transconductance was measured at $f_\mathrm{low}$, meaning for the lower top gate. The transconductance resembles the same characteristic resonances as the data in the first row. In particular, fluctuations with both slopes are seen. Again, we apply the conductance model to explain this observation. The total conductance contains contributions of the lower gate and the channel and can hence be expressed as
$\partial G_\mathrm{tot}/\partial V_\mathrm{TG,low}\approx\partial G_\mathrm{low}/\partial V_\mathrm{TG,low}+\partial G_\mathrm{constr}/\partial V_\mathrm{TG,low}$. Observing both slopes is therefore expected.

The third row shows data taken in a similar manner as for the second row, but the role of the upper and the lower top gate are reversed. As for the second row, we find that solely slope II resonances are observed in one configuration (see Fig.~\ref{c8fig4}~(f)) whereas both slopes are present in Fig.~\ref{c8fig4}~(e).

In summary, we have introduced a transport model for the split-gate structure investigated here, which assumes a parallel connection of three conductance channels. By tuning the gates individually, we could verify this model and attribute each set of resonances to one of the conductance channels. These findings support the above statement that the cause of the slope II oscillations is located in-between the gates.

\section{Conclusion}
Our measurements show that charge carriers can be guided to some extent through electrostatically defined nanoconstrictions. The observed interference effect, leading to fluctuations in the transconductance signal, was identified to originate from the narrow transport channel defined by the split-gates.

Recent experiments on dual-gated suspended bilayer graphene devices demonstrated, that phenomena like Coulomb blockade and quantized conductance are observable in gate defined quantum dots and quantum point contacts \cite{Allen:2012}. These free standing devices exhibit typically excellent transport quality, implying a smooth potential landscape. Alternatively, similar properties are achieved by using boron nitride as substrate \cite{Xue:2011}. Hence, bilayer graphene, which is encapsulated between two BN-flakes, is a promising candidate for electrostatically defined nanostructures in the ballistic transport regime \cite{Goossens:2012}.

\section*{Acknowledgement}
We acknowledge useful discussions and experimental support from C. R. Dean, A. F. Young and P. Kim. This research was supported by the Swiss National Science Foundation through the National Centre of Competence in Research 'Quantum Science and Technology'.
\newpage

\end{document}